\documentclass[10pt,journal,compsoc]{IEEEtran}
\usepackage{booktabs}
\usepackage{graphicx}
\usepackage{adjustbox}
\usepackage{authblk}
\usepackage{url}
\usepackage{multirow}
\usepackage[table,xcdraw]{xcolor}

\usepackage{tikz}
\def\checkmark{\tikz\fill[scale=0.4](0,.35) -- (.25,0) -- (1,.7) -- (.25,.15) -- cycle;}
\usepackage{pifont}

\newcommand{\xmark}{\text{\ding{55}}}
\usepackage{array,amsfonts,amsmath}

\begin{document}

\title{Artificial Intelligence for Emotion Recognition: Promise, Perils, and a Prospective Path for a Prosocial Future}

\title{Artificial Intelligence for Emotion Recognition: Promise, Perils, and Paving A Path Ensuring a Prosocial Future}

\title{Artificial Intelligence Based Emotion Recognition: Promise, Perils, \& a Prospective Prosocial Path Forward}

\title{AI-Based Emotion Recognition: Promise, Peril, and a Prospective Prosocial Path}





\title{AI-Based Automated Emotion Recognition (AER): Charting A Prosocial Future}

\title{AI-Based Automated Emotion Recognition (AER): Promise, Perils, and Prescriptions for Prosocial AER}

\title{AI-Based Automated Emotion Recognition: \\ Promise, Perils, and Prescriptions}

\title{AI-Based Emotion Recognition: Promise, Peril, and Prescriptions for Prosocial Path}







\author[1]{Siddique Latif\thanks{Email: siddique.latif@usq.edu.au}}
\author[2]{Hafiz Shehbaz Ali}
\author[3]{Muhammad Usama}
\author[1]{Rajib Rana}
\author[4,5]{Bj\"{o}rn Schuller}
\author[6]{Junaid Qadir}
\affil[1]{University of Southern Queensland, Australia}
\affil[2]{Emulation AI}
\affil[3]{National University of Computer and Emerging Sciences (NUCES), Pakistan}
\affil[4]{Imperial College London, UK}
\affil[5]{University of Augsburg, Germany}
\affil[6]{Qatar University, Doha, Qatar.}


\IEEEtitleabstractindextext{%
\begin{abstract}
Automated emotion recognition (AER) technology can detect humans' emotional states in real-time using facial expressions, voice attributes, text, body movements, and neurological signals and has a broad range of applications across many sectors. It helps businesses get a much deeper understanding of their customers, enables monitoring of individuals' moods in healthcare, education, or the automotive industry, and enables identification of violence and threat in forensics, to name a few. However, AER technology also risks using artificial intelligence (AI) to interpret sensitive human emotions. It can be used for economic and political power and against individual rights. Human emotions are highly personal, and users have justifiable concerns about privacy invasion, emotional manipulation, and bias. In this paper, we present the promises and perils of AER applications. We discuss the ethical challenges related to the data and AER systems and highlight the prescriptions for prosocial perspectives for future AER applications. We hope this work will help AI researchers and developers design prosocial AER applications.

\end{abstract}
}

\maketitle

\IEEEdisplaynontitleabstractindextext
\IEEEpeerreviewmaketitle
 
\section{Introduction}
\label{sec:Introduction}

Automated emotion recognition (AER) is an emerging multidisciplinary research area that leverages advances in artificial intelligence (AI) to algorithmically retrieve a person's emotional state using knowledge from psychology, linguistics, signal processing, and machine learning (ML). Development of AER capabilities can have a transformative effect on society with wide-ranging implications due to the critical role emotions play in human lives ranging from perception, learning, and decision-making \cite{barrett1987perspectives} \cite{zacharatos2014automatic} \cite{mlodinow2022emotional}. AER is an umbrella term that encompasses various related terms such as affective computing, affect recognition, emotional AI, or artificial emotional intelligence (AEI) that have been proposed in the literature for automated recognition of human emotions \cite{Schuller18-TAO}. 

\begin{figure}[!ht]
\centering
\includegraphics[width=0.5\textwidth]{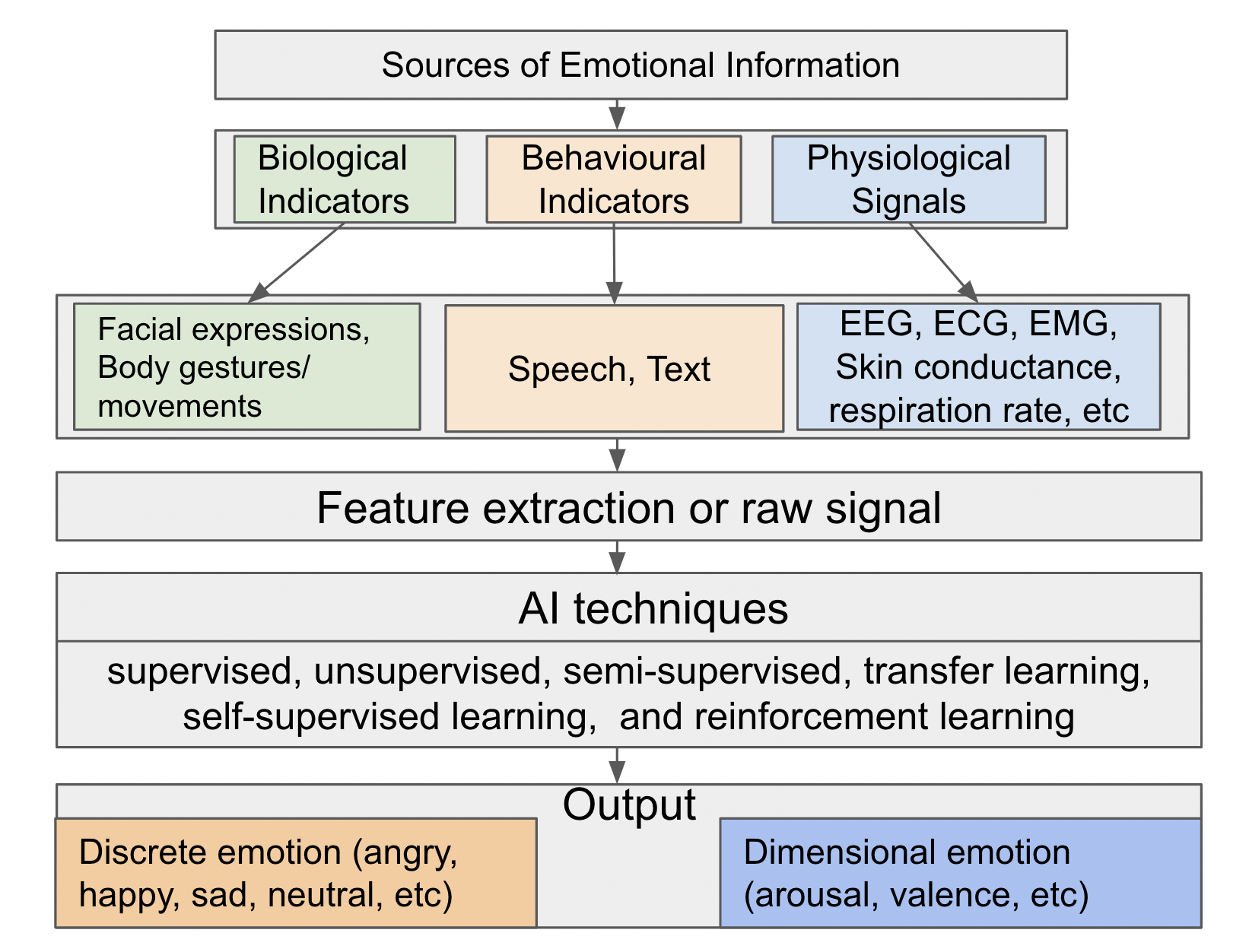}
\caption{An overview of AER systems, which can process emotional information from different emotional information sources to predict emotions by utilising different AI techniques.}
\label{fig:aersystem}
\end{figure}

In this paper, we use the term AER and focus on human-target AER. Human-targeted AER starts with active or passive sensors (e.g., a video camera, microphone, physiological sensor) that mainly captures contextual and behavioural data related to affective facial expressions, speech signals, body pose, gestures, gait, or physiological signals. It is imperative to utilise contextual information while identifying emotions \cite{latif2019direct}. The data obtained through the sensing devices extract emotional cues by the AI systems for \textit{categorical} or \textit{dimensional} emotion recognition as depicted in Figure \ref{fig:aersystem}. 


\begin{table*}[!ht]
\centering
\caption{comparison of our paper with related articles.}
\begin{tabular}{|l|lllll|}
\hline
 & \multicolumn{5}{c|}{Focus}   \\ \cline{2-5} 
\multirow{-2}{*}{Paper/Author (Year)}& \multicolumn{1}{l|}{Focus} & \multicolumn{1}{l|}{Promise} & \multicolumn{1}{l|}{Perils}   & \multicolumn{1}{l|}{Ethical Concerns}& Prescriptions \\ \hline

Kolakowska et al. \cite{kolakowska2014emotion} (2014) & \multicolumn{1}{l|}{AER} & \multicolumn{1}{l|}{\checkmark{}}  & \multicolumn{1}{l|}{\xmark{}}    & \multicolumn{1}{l|}{\xmark{}}                         &      \xmark{}         \\ \hline
Feng et al. \cite{feng2020review} (2020) & \multicolumn{1}{l|}{AER}& \multicolumn{1}{l|}{\xmark{}}                         & \multicolumn{1}{l|}{\checkmark{}}                         & \multicolumn{1}{l|}{\xmark{}}                         &        \xmark{}       \\ \hline

Batliner et al. \cite{batliner2020ethics} (2020) & \multicolumn{1}{l|} {Computational paralinguistics }& \multicolumn{1}{l|}{\xmark{}}       & \multicolumn{1}{l|}{\checkmark{}}                         & \multicolumn{1}{l|}{\checkmark{}}                         &        \xmark{}       \\ \hline

Liu et al. \cite{liu2011real} (2021)& \multicolumn{1}{l|}{AER (EEG only)}& \multicolumn{1}{l|}{\checkmark{}}  & \multicolumn{1}{l|}{\xmark{}} &\multicolumn{1}{l|}{\xmark{}}  &      \xmark{}         \\ \hline
 Mohammad et al. \cite{mohammad2022ethics} (2022)& \multicolumn{1}{l|}{AER}& \multicolumn{1}{l|}{\checkmark{}}  & \multicolumn{1}{l|}{\xmark{}} & \multicolumn{1}{l|}{\checkmark{}}                         &    \checkmark{}           \\ \hline

{\cellcolor[HTML]{EFEFEF}Our paper (2022)}& \multicolumn{1}{l|}{\cellcolor[HTML]{EFEFEF}AER} & \multicolumn{1}{l|}{\cellcolor[HTML]{EFEFEF}\checkmark{}} & \multicolumn{1}{l|}{\cellcolor[HTML]{EFEFEF}\checkmark{}} & \multicolumn{1}{l|}{\cellcolor[HTML]{EFEFEF}\checkmark{}} & \multicolumn{1}{l|}{\cellcolor[HTML]{EFEFEF}\checkmark{}}  \\ \hline
\end{tabular}
\label{tab:com}
\end{table*}

AER has evolved over the years and achieved remarkable advances; however, it faces various complex and critical challenges that escalate the need for further research to design more trustful and beneficial systems \cite{gunes2011emotion}. Some major challenges faced in AER are: 

\begin{enumerate}
\item Unavailability of large datasets, which restricts the exploitation of powerful DL models to achieve better performance \cite{latif2021survey}; 
\item Collection of real-life data for modalities such as brain activity and neurotransmitters is very challenging; 
\item Varied idiosyncratic nature of human emotions due to which it is hard to accurately recognise them; 
\item Judging varying emotions in real-time is hard as most vision and speech-based AER algorithms focus on identifying the peak high-intensity expression by ignoring lower-intensity ones, which can result in inaccuracies; and 
\item Cultural differences in manifesting emotions, which makes the problem of developing global AER very difficult. 
\end{enumerate}

The public use of AER services also raises multiple privacy and security-related concerns due to the intimate nature that the AER systems detect, process, recognise, and communicate. This paper presents promising applications of AER and discusses its various perils. In particular, we focus on presenting the ethical concerns related to AER systems and databases to highlight the prescriptions for future AER prosocial systems. 

We note here the specificity or universality of human emotions has been a long-standing debate \cite{darwin2015expression}. The proponents of the universality of emotions suggest that emotions can be recognised regardless of the different cultural backgrounds. While theoretical studies \cite{ekman1994strong,ekman1992there} on multicultural studies have suggested six basic universal emotions, current AER systems do not perform well in multicultural settings. 

The novelty and contributions of our paper are highlighted in Table \ref{tab:com}, where we compare this paper with the existing articles on AER. The article by Mohammad et al. \cite{mohammad2022ethics} enlists the ethical challenges for AER and suggests future directions but does not cover AER applications or the challenges related to bias, adversarial attacks, explainability, etc. Similarly, other articles only focus on modality-specific applications \cite{liu2011real} or general challenges \cite{feng2020review} without focusing on ethical issues. In this paper, we attempt to present AER's promise, perils, and ethical concerns. We also provide prescriptions for designing prosocial AER systems. We hope this paper will guide navigating research and ethical implementation choices for anyone who wants to build or use AER for research or commercial purposes.

The rest of the paper is organised as follows. The promise of AER is described in Section \ref{sec:Promise}. The various perils of AERs are detailed in Section \ref{section:perils}. The major ethical concerns with AER are discussed in Section \ref{sec:ethical}. We elucidate a prospective path to a prosocial future for AER in Section \ref{sec:prosocial}. The paper is concluded in Section \ref{sec:Conclusions}.

\section{Promise of AER}
\label{sec:Promise}

AER has a wide range of applications in fields such as healthcare, entertainment, advertising, customer service, transportation, employment decisions, tutoring systems, law enforcement, and human-computer interaction. Applications of AER are classified according to the input signal provided in Table \ref{applications}. A broad description of frequently targeted AER applications in different domains is presented in Table \ref{table:promise} and briefly discussed below.
\begin{table}[!th]
\centering
\caption{Applications of AER and input data.}
\scriptsize
\begin{tabular}{|l|l|l|}
\hline
\textbf{\begin{tabular}[c]{@{}l@{}}Human \\ expression\end{tabular}}  & \textbf{Data}                                                              & \textbf{\begin{tabular}[c]{@{}l@{}}Possible applications\end{tabular}}                                                                                   \\ \hline
\begin{tabular}[c]{@{}l@{}}Vocal\\ expressions\end{tabular}           & Audio                                                                      & \begin{tabular}[c]{@{}l@{}}Call centres, meetings, voice \\ assistants, social robots, educations,\\  human resource, healthcare, etc.\end{tabular}        \\ \hline
\begin{tabular}[c]{@{}l@{}}Facial \\ expressions\end{tabular}         & Visual                                                                     & \begin{tabular}[c]{@{}l@{}}Autonomous vehicles, industrial\\  and social robots, surveillance, \\social media, gaming, \\education, healthcare.\end{tabular} \\ \hline
\begin{tabular}[c]{@{}l@{}}Body movements \\ and posture\end{tabular} & Visual                                                                     & Surveillance, education, healthcare.                                                                                                                      \\ \hline
\begin{tabular}[c]{@{}l@{}}Physiological \\ signals\end{tabular}      & \begin{tabular}[c]{@{}l@{}}EEG and ECG\\  records, heart rate\end{tabular} &\begin{tabular}[c]{@{}l@{}} Wearable devices and \\medical equipment\end{tabular} .                                                                                                                     \\ \hline
\end{tabular}
\label{applications}
\end{table}

\subsection{Healthcare}
Developing AER methods in healthcare can greatly enhance the quality of life, enable individuals to better understand and control their affective states (e.g., fear, happiness, loneliness, anger, interest, and alertness), and mitigate various psychological issues (that could have resulted in incidents of suicide, homicide, disease, and accident) \cite{latif2020speech}. This can greatly improve quality of life and help achieve long-term goals \cite{darwin2015expression}. It can help save many lives by monitoring people and regulating their emotions through stressful times (e.g., in pandemics or economic crises). It also minimises counter-productive behaviour such as suicidal tendencies and or anti-social behaviour. In healthcare settings, AER services play a pivotal role in shaping the healthcare functionality and communication among professionals, thereby improving professional-patient relations \cite{blanch2012effective}. It can help design assessment and monitoring of emotional consequences due to different illnesses. For example, AER systems can be potentially used to monitor the patient-physician relationship in chronic diseases \cite{brezulianu2022not}. This will help improve the management of chronic illnesses.

The importance of AER technology has also come to the fore amid the ongoing global economic and public health crisis during the COVID-19 pandemic \cite{zhang2020emotion}. The pandemic situation impacts people physically, mentally, and economically. AER systems can help to analyse and understand emotional responses during such crises affecting mental health. Studies \cite{kabir2021emocov,imran2020cross,al2021monitoring} show that the negative emotions among the population increase during the pandemic, i.e., COVID-19, and people become optimistic over time by adapting to the pandemic. In global crisis situations like COVID-19, AER systems can help measure cross-cultural emotional trends to learn the correlation among populations despite the socio-economic and cultural differences \cite{imran2020cross}.


\begin{table}[!t]
\scriptsize
\centering
\caption{Summary of promises of AER technology in different domains. }
\begin{tabular}{|l|l|}
\hline
Domains                           & Promise                                                                                                                                                                                                         \\ \hline
Healthcare                        & \begin{tabular}[c]{@{}l@{}}Monitoring people and regulating emotion.\\ Improves patient-physician relationships.\\ Analyses and understands emotions in \\natural disasters and crises.\end{tabular}            \\ \hline
Education                         & \begin{tabular}[c]{@{}l@{}}Improves student-teacher interaction.\\ Quantifies student moods and engagement \\ in the classroom. \\ Promotes effective learning and increase\\  students’ interest.\end{tabular} \\ \hline
Safety                            & \begin{tabular}[c]{@{}l@{}}Improves workplace safety. \\ Enables help for emotionally suffered \\ co-workers.\\ Monitors the drivers’ fatigue, stress, etc.\end{tabular}                                        \\ \hline
\begin{tabular}[c]{@{}l@{}}Law enforcement\\ and forensics \end{tabular}    & \begin{tabular}[c]{@{}l@{}}Helps identify threats of violence \\ and terrorism.\\ Provides additional aid in criminal\\ investigation.\end{tabular}                                                             \\ \hline
Advertising and Retail            & \begin{tabular}[c]{@{}l@{}}Helps maximise customers' engagement. \\ Helps retailers to make decisions on \\ product pricing, packaging, branding, etc. \\ Helps improve advertisement strategies.\end{tabular}  \\ \hline
\begin{tabular}[c]{@{}l@{}}Emotional and Social\\ Intelligence\end{tabular} & \begin{tabular}[c]{@{}l@{}}Helps influence the mood of the \\ overall population.\\ Helps leaders and decision-makers to \\ handle highly challenging situations.\end{tabular}                                  \\ \hline
Monitoring and Evaluation         & \begin{tabular}[c]{@{}l@{}}Enables monitoring of employees\\ performance. \\ Enables monitoring of  major psychiatric\\ problems in both military and civilian.\end{tabular}                                    \\ \hline
Gaming                            & \begin{tabular}[c]{@{}l@{}}Monitors players’ emotional states \\ and dynamics during gameplay. \\ Helps design affect-aware video games.\end{tabular}                                                           \\ \hline
\end{tabular}
\label{table:promise}
\end{table}

\subsection{Education}
Emotions are very crucial in the education systems due to their important role in the cognitive processes responsible for assimilating and learning new
information \cite{bouhlal2020emotions}. Unfortunately, the current education system fails to track students' emotions and hidden indicators of their internal feelings, thus, making it delicate to adapt and keep the communication channel intact. It has been found that the identification and monitoring of the learner's emotional state greatly facilitates the teacher in taking actions that significantly enhance the tutoring quality and execution and improve student-teacher interactions \cite{bulut2018effects}. Therefore, it is worthwhile to utilise smart systems that can model the relations between emotions, cognition, and action in the learning context \cite{pritchard2003using}. In this regard, AER systems can be considered by schools to quantify student moods and engagement in the classroom \cite{denervaud2020emotion}. AER could help to reinforce students' attention, motivation, and self-regulation toward studies. It could also help promote effective learning by increasing students' interest \cite{pritchard2003using}. On the other hand, AER systems can improve certain emotional qualities teachers must have to facilitate pedagogical approaches in education.

\subsection{Safety}
Emotions are directly linked to human problem-solving abilities \cite{clore2007emotions}. Safety behaviours can be predicted from the individuals' ability to manage and process emotions during a time of stress. There is ample evidence that negative emotions such as anger, fear, and anxiety strongly affect human behaviour and occupational safety \cite{zhang2021effect}. For example, emotions can impact workplace safety and health. In the workplace, the negative mood of a person can contaminate an entire team or group. This may damage workplace safety and impair team performance. If such behaviours are left unaddressed, negative emotions can be a workplace hazard, with visible effects on team safety. AER systems can provide better solutions to monitor an individual's mood and emotions. In addition, these systems can help find workers who might need help.  

In transport, AER systems can be utilised to improve the safety of drivers as well as anyone on the road. Driving occupies a large portion of our daily life and is often associated with the cognitive load that can trigger emotions like anger or stress, which can badly impact human health and road safety \cite{zepf2020driver,ding2014driving}. Studies \cite{davoli2020driver,jeon2014effects} show that induced negative emotions like anger can decrease a driver's perceived safety and performance compared to neutral and fearful emotional states. AER services are being utilised in automotive 
environment 
to monitor the drivers' fatigue, alertness, attention, and stress level to improve automotive and industrial safety by avoiding serious accidents \cite{zepf2020driver}. Mass adoption of AER systems to monitor psychological and physiological parameters can significantly enhance the detection of dangerous situations. 
    
\subsection{Law enforcement and forensics}

AER systems are increasingly being used for law enforcement, and forensics, where such systems have many possible applications in identifying threats of violence and terrorism and detecting lies and fraud \cite{hollien2002forensic}. In a forensic investigation, a lie can arise from denial, evasion, distortion, outright fabrication, and concealment by offenders to appear non-accountable for their exertions \cite{roberts2012forensic}. AER systems can help law enforcement agencies to detect deception or malingering by identifying reliable emotional clues. In this way, AER systems provide additional aid and insights to law-enforcement agencies while pursuing criminal investigations \cite{lawenf}. AER systems can also help detect and differentiate between acted and genuine victims \cite{roberts2012forensic}.

    
    
\subsection{Advertising and Retail}
In marketing, one of the best ways to sell products is to engage the customers emotionally. Companies employ vast resources for affective marketing by maximising user engagement with AI. They attempt to understand and appeal to the customers' interests, and emotions \cite{alajmi2013shopmobia}. In order to gauge a shopper's emotion, AER systems use sensing devices installed in high-traffic locations, including entrances, aisles, checkouts, etc. AER systems detect the emotional responses of individual shoppers, which help retailers in making decisions on crucial factors, including product pricing, packaging, branding, or shelf placement. In this way, AER systems help retailers understand how consumers communicate both verbally and non-verbally, which may help fuel customers' buying decisions.

Emotions highly impact individuals' responses to receiving marketing messages. Therefore, sending an emotionally tailored message to the target audience increases the customers' attention to the advertisement. This helps companies to increase the product's appeal and achieve a higher level of brand recall \cite{otamendi2020emotional}. Indeed, advertisements with emotional content have more potential to be remembered than those conveying notification \cite{page1990memory}. 


\subsection{Emotional and Social Intelligence}
Emotional and social intelligence involves 
understanding inside oneself, 
observing, and interpreting others for cognitive and emotional empathy, and responding constructively in a given situation. There is great interest in politics to capture and influence the mood of the overall population or community to understand patterns of emotional contagion \cite{lievens2017practical}. Emotional and social intelligence can help leaders and decision-makers pick up emotional cues from a population and handle highly challenging situations. Social networks are particularly utilised to understand population behaviours \cite{krause2007social}. 

\subsection{Monitoring and Evaluation}
AER systems are being utilised to screen candidates in interviews \cite{su2021predicting} and to evaluate and monitor employees' fatigue, stress, happiness, and job performance \cite{momm2015pays}. It is widely accepted that emotional intelligence directly influences an employee's intellectual capital, organisational reactivity and 
retentively, production, employee appeal and ability to provide good customer service \cite{subhashini2015analyzing}. AER systems can 
contribute to 
assess a candidate's suitability for a job and measure important traits like dependability and cognitive abilities. In particular, embedded AER systems enabled through IoT can provide fine-grained analysis of emotions and sentiments \cite{jarwar2017exploiting}, which can be used in various ways for monitoring and evaluations. In the military and other defence-related departments, AER systems are 
partially 
used to track how sets of people or countries `feel' about a government or other entities \cite{mohammad2022ethics}.

\subsection{Gaming}
Video games are related to the burgeoning area of entertainment applications. Millions of users across the globe are entertained by violent games \cite{teng2019longitudinal}, and most selling games contain violence and aggression. These video games are played by adolescents \cite{coyne2020pathological}. For instance, in the United States, 81\% of adolescents have access to digital games, and on average, a gamer spends 6 to 8 hours a week playing video games \cite{miedzobrodzka2021insensitive}. AER systems are highly suited to be utilised for the design of affect-aware gaming platforms that can monitor players' emotional states and dynamically change the game's theme to more effectively engage the player \cite{szwoch2015emotion}. In these ways, affect-aware video games with an entertainment character can also be utilised to initiate pro-social behaviour by preventing anti-social actions along with various applications such as e-learning, marketing systems, and psychological training or therapy.

\section{Perils of AER}
\label{section:perils}

AER technology has a wide range of potentially intrusive applications, as discussed in the previous Section \ref{sec:Promise}. It uses biometric data that may be used to reveal private information about individuals' physical or mental states, feelings, and thoughts. It can also be used to interfere with the formation of beliefs, ideas, and opinions. Modern AER systems often use deep learning (DL) models to obtain state-of-the-art performance. However, such DL models are known to be inscrutable and are also not robust and are vulnerable to bias and poor performance in the face of distribution shifts and adversarial attacks \cite{pitropakis2019taxonomy}. This raises concerns about using the validity of AER services since it is not uncommon to see that even well-intentioned technologies can have negative consequences and how technologies can end up harming rather than benefitting people \cite{latif2019caveat, o2016weapons}. We discuss some of the prominent perils of AER next.

\subsection{Risk of Exploitative Manipulation} AER technology can be exploited and used to influence and control driving markets, politics, and violence. Already, there is a big concern in the community about 
major technology 
companies morphing into empires of behaviour modification \cite{russell2019human}. With AER having access to intimate human emotions, the risk of exploitative manipulation rises further as such information can be used to interfere with the formation of beliefs, ideas, opinions, and identity by influencing emotions or feelings \cite{greene2020ethics}.

\subsection{Lack of Consent and Privacy Violations} AER systems utilise AI technology in their design with biometrics or other kinds of personal data (speech, facial image, among others). This allows for information about physical or mental health, thoughts, or feelings---which an individual may not want to choose to share---to be automatically inferred without the person's consent or knowledge. This has grave privacy concerns and can be used to establish and strengthen oppressive dystopian societies.

\subsection{Lack of Explainability/Accountability}
AER systems usually lack explainability due to the complex internal mechanics of the AI model and the wide-scale adoption of BlackBox models based on ``deep learning'' technology. This inability to understand how AI performs in AER systems hinders its deployment in law, healthcare, and enterprises from handling sensitive consumer data. Understanding how AER data is handled and how AI has reached a particular decision is even more critical for data protection regulation. Explainability of AER services will allow companies to track AI decisions and monitor biased actions. This will also ensure that AER processes align with regulation and that decision-making is more systematic and accountable.  

\subsection{Vulnerability to Adversarial Attacks} Modern AER-based tools typically rely on ``deep learning'' based models such as those built on deep neural networks (DNNs), which are composed of multiple hidden layers. DNNs are also quite fragile to very small specially-crafted adversarial perturbations to their inputs. This can cause false prediction in AER systems \cite{latif2020deep}, which might have adverse consequences. For instance, an adversarially crafted example can cause an AER system to diagnose mental diseases inaccurately. This is one of the critical concerns of integrating AI-based services like AER in real-life.

\subsection{Vulnerability to Bias} There is scepticism in the community regarding the efficacy of AER and fears that using AER may accentuate and institutionalise bias \cite{purdy2019risks}. Since getting accurately labelled data is very expensive and time-consuming, any embedded bias in large annotated emotional training data is likely to be built into any systems developed using such data. Most of the AER systems use laboratory-designed datasets based on actors simulating emotional states in front of a camera. Furthermore, the labels used by ML models typically represent perceived emotion rather than felt emotion since the majority of the existing AER datasets are labelled by human annotators based on their perception \cite{greene2020ethics}. For instance, in facial emotion recognition, the labels for a photograph are provided by annotators, not by the individual in the photographs \cite{greene2020ethics}. This might not represent genuine inner emotions and may contain hidden biases that may become apparent only after deployment.

\subsection{Reductionist Emotional Models} AER algorithms base their working on basic emotion theories \cite{ekman1971constants} that have been widely critiqued \cite{leys2011turn}. For instance, the widely applied theory posited by Paul Ekman regarding six universal emotions (happiness, sadness, fear, anger, surprise, and disgust) that can be recognised across cultures from facial expressions has been criticised by experts as being too reductionist \cite{crawford2021atlas}. An automatic link between facial movements and emotions is assumed---e.\,g., a smile means someone is happy. However, this might not always be true. For instance, in the US and many other parts of the world \cite{tsai2016leaders}, it is common to smile at strangers, which might not represent inner feelings or states. It follows that more contextual details are required to understand the emotion, potentially requiring more data and invasive practices.

\subsection{AI's White Guy Problem or \textit{Neo-Colonialism}}
Some findings indicate that AI technology suffers from problems such as sexism, racism, and other forms of discrimination \cite{hong2020sexist}. A major aspect related to this arises from homogeneous or unrepresentative data. Another reason could be focusing on the majority class since optimising for the majority class will
usually 
improve overall accuracy. Unfortunately, this translates into discrimination against the minority classes as AI models typically do not automatically provide fairness unless constraints are placed for ensuring fairness (in which case, the overall accuracy will 
usually 
reduce as fairness and accuracy are different objectives, and it is not uncommon for them to have tradeoffs) \cite{sahbaz2019artificial}. If we do not work to make AI more inclusive, we risk creating machine intelligence that ``mirrors a narrow and privileged vision of society, with its old, familiar biases and stereotypes'' (Kate Crawford, New York Times, \url{https://tinyurl.com/2h6fu8dv)}). Experts are now calling out for using decolonial theory as a tool for sociotechnical foresight in AI to ensure that the hegemony resulting from the domination of the AI industry by a limited number of demographic groups and nations does not have harmful effects globally \cite{mohamed2020decolonial}.

\section{Ethical Concerns with AER}
\label{sec:ethical}

\begin{figure*}[!t]
\centering
\includegraphics[width=0.7\textwidth]{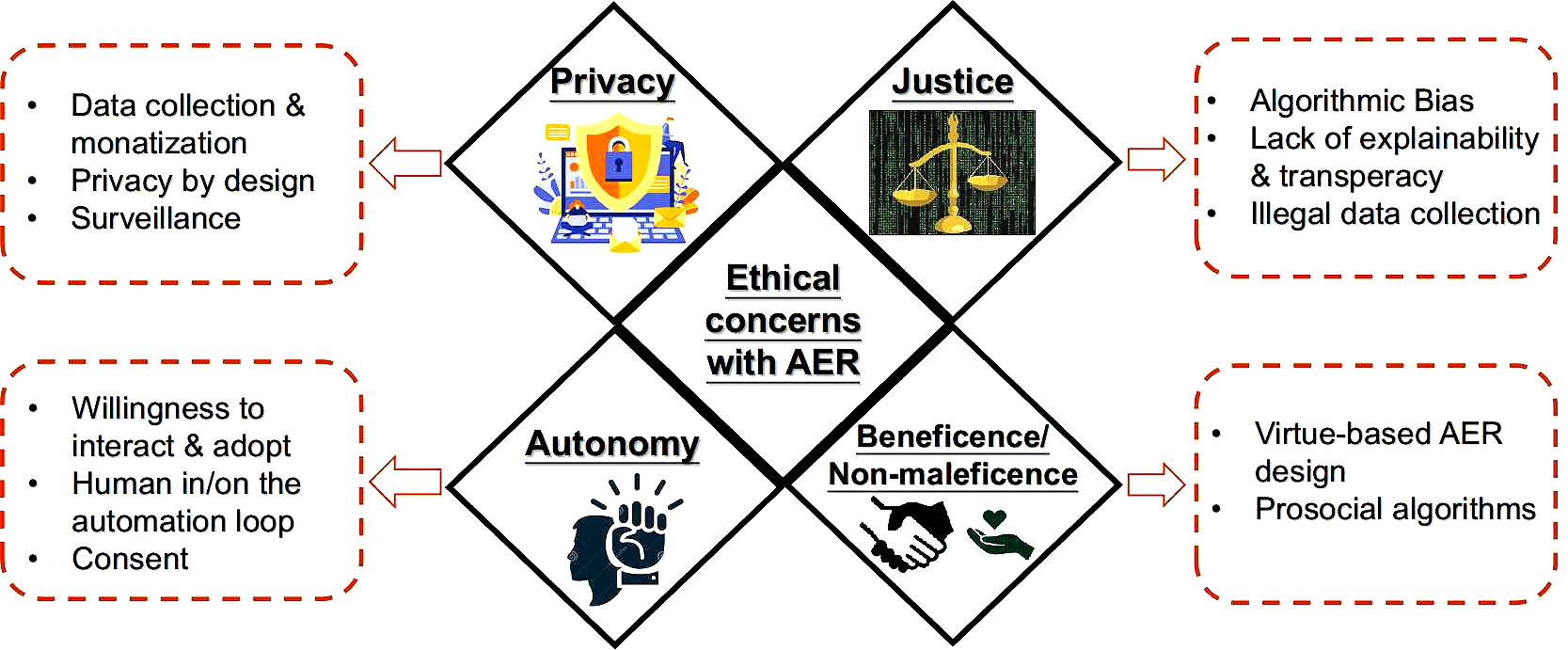}
\caption{Summary of ethical concerns associated with AER. }
\label{fig:ethic}
\end{figure*}

Giving emotions to a computer is another term for AER technology \cite{picard2000affective}. It is exciting and a pipe dream to have a human-like or superior emotion detection system. 
In the last decade, techniques based on advanced techniques in ML and deep learning have outperformed almost all classical methods in recognising and understanding human emotions from facial, speech, and text inputs. These advanced learning techniques have produced effective and efficient results in AER and automated the whole process. 
AER systems are used in the commercial market for understanding user engagement, sentiment analysis, attention tracking, behaviour understanding, etc. However, these AER systems are also prone to shortcomings and biases in the training and testing data. 
The literature on the shortcomings of traditional and deep ML techniques suggests that data and algorithmic biases can impact the performance of these learning techniques \cite{crawford2021atlas}. AER systems are developed using data acquired from humans, and human biases are likely to be translated into the learning process, impairing AER system judgements \cite{dubber2020oxford}. There is a need to enforce responsible AI practices \cite{domnich2021responsible}, and ethical guidelines for the design, development, and integration of AER systems in the wild \cite{shaw2019artificial}.   

The use of AER for emotional surveillance raises many ethical concerns, which motivates the need to identify basic ethical principles and guidelines that address ethical issues arising from the use of AER technology on human subjects to ensure that human subjects are not exploited or manipulated. In this regard, we can look at a traditional consensus on basic principles such as those expressed in the Belmont Report produced in 1979 by the US National Commission for the Protection of Human Subjects of Biomedical and Behavioural Research (\url{https://tinyurl.com/5pr5rpe9}). The Belmont Report identified three main principles---(1) \textit{respect for persons}; (2) \textit{beneficence}; and (3) \textit{justice}---in their study focused on documenting the basic ethical principles and guidelines that should direct the conduct of biomedical and behavioural research involving human subjects. In light of the described perils of uncritical use of technology and the various ethical and moral dilemmas posed by AI \cite{denning2020dilemmas}, a lot of attention has focused on incorporating ethics in the field of AI leading to a proliferation of AI ethics principles and code of ethics. Interestingly, Jobin et al.\ \cite{jobin2019global} have highlighted 84 such codes of ethics related to AI in 2019 and found that four high-level ethical principles---\textit{beneficence}, \textit{non-maleficence}, \textit{autonomy}, and \textit{justice}---capture the essence of most AI declarations with Floridi and Cowls \cite{Floridi2019Unified} also adding \textit{explicability} as a high-level principle demanding that AI models should not work as inscrutable blackboxes. We summarise the AER-related ethical concerns in Figure \ref{fig:ethic} and discuss these concerns in detail next.

\subsection{Ethical Concerns Related to Justice}
AI is being used in every facet of daily life, including criminal justice, social media, social justice, health care, smart cities, and urban computing. Although it has been well stated in the literature that AI-based systems are incapable of understanding the concepts of justice and social standards \cite{surden2021machine}. Buolamwini et al.\ \cite{buolamwini2018gender} emphasise that the AI-based facial detection system discriminates against gender and people of colour. They also demonstrated that commercial AI-based facial detection systems need a firmer grasp of ethics and auditing \cite{raji2019actionable}. Cathy O'Neil et al.\ \cite{o2016weapons} exposed the flaws in employing big data and AI-based algorithms to make choices with real-life consequences, and these consequences are leading to a societal split and shattered democracy. The ethics of applying AI in law and its obstacles are discussed in \cite{surden2020ethics, surden2019artificial, surden2014machine, surden2022values}. In contrast, the ethical issues of employing AER systems for learning expressions and privacy concerns are detailed in \cite{mcstay2020emotional}. Wright \cite{wright2021suspect} explains the opacity of algorithms employed in AER systems, the inadequacy of AI in comprehending human emotions, and how these failings lead to an unjust society \cite{crawford2021artificial}. Carrillo\ \cite{carrillo2020artificial} discusses the ethical AI debate from the standpoint of law and how AI shortcomings impede the general application of the AI-based judicial system. Finally, Khan et al.\ \cite{khan2021ai} present a thorough discussion of AI-enabled face recognition systems and their ethical implications in the criminal justice system.

In the last few years, AI-based predictive policing tools are becoming a part of global criminal justice systems. These systems are largely based on facial recognition technology with added emotion recognition and DNA matching. These tools have many ethical issues \cite{o2016weapons, buolamwini2018gender}. Millerai \ \cite{millerai} provides a comprehensive discussion on the ethical issues of predictive policing and facial recognition systems in criminal justice systems. They argued that these systems violate privacy rights, autonomy rights, and basic human morality. They also discuss the misuse of AI-based predictive policing and facial/emotion recognition tools in liberal democracies and the dangers of similar technology in authoritarian states. In order to use 
With AI-based systems making critical judgements about individuals (hiring process, advertising process, etc.), it is vital to consider and address ethical concerns. Automated physiognomy refers to the use of AI models to identify a person's gender, emotional state, level of intellect, and other characteristics from only one photograph of them. 
Engelmann et al.\ \cite{engelmann2022people} debated the fairness and ethical concerns of automated physiognomy with a comprehensive experiment in which thousands of non-AI individuals were invited to respond on what AI should ethically infer from faces. The questions also include the number of characteristics inferred from faces by well-known AI models (including AER models), such as gender, emotional expression, likeability, assertiveness, intellect, colour, trustworthiness, and use of spectacles. Because all these characteristics are subjective, participants were asked to provide a Likert scale score and a written explanation of why a particular score was awarded for two specific use cases: advertising and hiring \cite{engelmann2022people}. 
The overall findings show that individuals, independent of context, substantially disagree with the automated physiognomy regarding assertiveness, likeability, trustworthiness, and intellect. Participants were also observed to be more dissatisfied (ethically) with the AI inferences about race, gender, emotional expression, and wearing spectacles in the hiring use case \cite{engelmann2022people}. AER systems suffer from the same issue, and the results reveal that a lack of auditing will result in an unfair automated judgement, which will have far-reaching effects on the social justice system.

Podoletz \cite{podoletz2022we} investigated the use of emotional AI (a blend of affective computing \cite{picard2000affective} and AI that gives probabilistic predictions of a person's or community's emotional state based on data points about the individual or community) in criminology, police, and surveillance. Given the ethical concerns, algorithmic biases, and annotation issues, Podoletz urge that emotional AI not be implemented in public spaces since these technologies will expand policing authority, raise privacy concerns, and operate as an oppressive instrument in authoritarian states. Podoletz goes on to claim that deploying emotional AI tools like AER would result in a highly regulated and controlled society, causing a severe schism in the social justice system. Lastly, \cite{podoletz2022we} discusses the repercussions of using emotional AI tools in crime predictions and preemptive deception detection. Minsky \cite{minsky2007emotion} in his famous book ``The emotion machine: commonsense thinking, artificial intelligence, and the future of the human mind'' talked about emotional AI and its relation to basic cognition and neuroscience. He also talks about the ethical challenges in AI systems designed for emotion recognition. Emotional AI (affective computing paired with AI) technologies are used for reading, interpreting, replicating, and influencing human behaviour and sentiments, according to Yonck \cite{yonck2020heart} in his book ``Heart of the machine: our future in a world of artificial intelligence." The author also discusses the moral dilemmas raised by the commercial application of these technologies. He further  contends that the code of ethics designed for emotional AI tools like AER systems would be subverted in markets in favour of monetary and political gains, thus undermining the sociopolitical justice of society \cite{yonck2020heart}. Van \cite{van2020ethical} elaborated upon the ethical issues in AI-based facial recognition technologies (face, gender, class, race classification, AER systems, and others). The report demonstrates how one could use face recognition technology as an instrument of oppression, with a huge surveillance engine created to monitor and classify minorities and, by extension, a whole country. 


The AER sector is predicted to be worth \$26 billion by 2026 \cite{crawford2021time}. Crawford et al. \cite{crawford2021time} recommend that AER systems be regulated as soon as possible. She claims that several technology businesses used the pandemic as a justification to introduce emotion detection systems to assess the emotional state of employees and even children. She presented the example of an AER system called 4 Little Things\footnote{\url{https://www.4littletrees.com/}}, which is used to infer children's emotions while carrying out their classwork with no supervision or regulation. She also states that with AER systems now being widely employed in many socioeconomic areas (hiring, healthcare, education, advertising, among others), it is important that this industry be regulated to avoid injustice and the fostering of an unjust society. A report on the ethical issues related to biometric applications (including AER) in public settings was published by the Citizens' Biometrics Council \cite{ada2021citizens}. The suggestions are based on conversations in public concerning the ethics of using AER and other biometric technology. The report urges the establishment of a comprehensive regulatory framework for biometric systems, a credible oversight agency, and minimum standards for designing and deploying face and AER systems \cite{ada2021citizens}.

As previously described, it has been observed in the literature that AI models do not automatically provide fairness or justice unless it is explicitly asked for \cite{kearns2019ethical}. As Stuart Russell describes in his book, a problem underlying the model of conventional optimisation-based AI is that you only get what you explicitly ask for with the unstated assumption being that you implicitly agree that you do not care at all about everything you do not specify \cite{russell2010artificial}. The author calls this the King Midas problem of AI referring to the Greek mythological story in  which King Midas gets all that he specifies, but the situation still ends unacceptably since he did not specify exactly what he did not want (and unacceptable values were incorrectly inferred) \cite{russell2010artificial}. Various studies have shown that AER technology is prone to bias and can suffer from a lack of fairness, accountability, and transparency. 
This has real consequences when such technology is used for critical decisions, such as in judicial systems for making judgements about sentencing \cite{o2016weapons,crawford2021atlas}. Therefore, AER technology requires a continued and concerted effort to address such issues, because misreading an individual's emotions can cause severe consequences in specific scenarios. 

\subsection{Ethical Concerns Related to Beneficence/ Non-maleficence}
Ethical principles of beneficence (``do only good")  and non-maleficence (``do no harm") are closely related. Beneficence encourages the creation of AI services to promote the well-being of humanity and the planet, while non-maleficence concerns the negative consequences of AI \cite{sft}. These concerns are also important in the designing and deployment of AER technology. Therefore, AER services should avoid causing both foreseeable and unintentional harm. This requires a complete understanding of AER technology and its scientific limitations to manage the attendant risks. The services should be designed to benefit human beings and increase their well-being to make AER prosocial.

Designing a Prosocial AER system requires mitigation of ethical concerns highlighted in the literature \cite{wright2021suspect}. With the unprecedented penetration of social media applications and the use of surveillance technologies, the opt-in and opt-out model of data sharing is long gone. Now, most of the applications gather data irrespective of permissions, and the written conditions that one agrees to upon usage are written in a language that is a challenge for the regular user. It is problematic, and many incidents of unethical use of the data are being reported in the literature. Unfortunately, the idea of beneficence / non-maleficence is not considered as vital as it should have been in designing AER systems.

Beneficence / non-maleficence principles are based on moral conscientiousness, social good, and trustworthiness of people, companies, and algorithms. Raquib et al. \cite{raquib2022islamic} propose a virtue-based ethical design of AI systems, although the debate is philosophical and many areas of the subject suffer from a lack of generality. The topic of virtue-based ethical systems and the ethical quandaries raised are also pertinent to AER systems. Because AER systems are meant to learn from user behaviour and how that behaviour may be watched, hugged, and altered, the essential nature of the data and the influence of the AER system on society necessitates an AER design that is founded on beneficence / non-maleficence. Examination supervision technologies have saturated the market under the guise of COVID-19. These tools are often AI-based, with face and emotion recognition algorithms used to monitor exam participants. Though these methods are intended to assure that the examination is conducted correctly, they lack core ethical standards such as privacy, transparency, fairness, and beneficence. Coghlan et al. \cite{coghlan2021good} examined and reported ethical challenges with AI-based examination supervision tools, arguing that the issues will not be resolved until ethical standards are not included in the basic design principles of AI-based automated systems like facial recognition and AER. Similarly, the reality of social robots is just around the horizon, and numerous AER-based robots are presently being employed in a variety of social contexts, and the number of these robots is rapidly increasing. The ethical challenges raised by social robots originate from the fundamental debate about the uncertainty and responsibility of AI systems. Bosch et al. \cite{bochrobotic} provide a brief description of the ethical risks involved in social robotics, including how the concept of doing only good and not harm is required for social robots, as well as various technological and social challenges associated with developing such ethics in robots.

\subsection{Ethical Concerns Related to Privacy}
AER services mostly use DL algorithms that are trained on masses of data to learn and perform decision-making. Ethical concerns related to privacy require protecting individuals' data and preserving their privacy. Over the last two decades, the rise of surveillance capitalism went largely unchallenged. Tech companies like Google and Facebook provide free online services and use personal data for mass surveillance over the internet. Such companies collect and scrutinise users' online behaviours including searches, purchases, likes, dislikes, and more, to predict, modify, and control users' behaviours. Lanier has coined the term BUMMER\footnote{\url{https://www.theguardian.com/technology/2018/may/27/jaron-lanier-six-reasons-why-social-media-is-a-bummer}}, or ``Behaviours of Users Modified, and Made into an Empire for Rent", for the economic model followed by big tech corporations in the world of surveillance capitalism.

The design of AER systems depends heavily on face recognition technologies, and it is advised in the literature that emotion recognition systems should be regularly updated and audited \cite{crawford2021time}. Bowyer \cite{bowyer2004face} discuss facial recognition systems' security vs the privacy dilemma. The right to privacy is a fundamental right guaranteed by practically every country's constitution. Many countries use facial recognition systems, and by extension AER systems, for mass surveillance without the agreement or scrutiny of regulatory organisations, which is a serious concern in the domain of technology's social effect. Bowyer \cite{bowyer2004face} argues that utilising these recognition technologies violates the constitutionally guaranteed right to privacy. AER systems not only employ facial recognition technologies but also infer the emotional state and other aspects of the face without consent, which is an abuse of power and a blatant violation of the fundamental right to privacy. The effectiveness of a security surveillance system is determined by the performance of the facial recognition system and a combination of the algorithms to measure the underlying emotional states and motives from just an image of the face and body. These algorithms and facial recognition systems have shown to be biased and unreliable in the literature \cite{stark2018facial}; false positives and negatives have life-threatening repercussions, and privacy infringement concerns are unprecedented \cite{bullington2005affective}.

The discussion of privacy and the right to privacy has become prevalent with the advent and adoption of new technological applications such as internet-of-things (IoT), robotics, pervasive technologies, biometric technologies, augmented and virtual reality, and digital platforms \cite{royakkers2018societal}. AER systems are used in homes, health care facilities, childcare centres, social media apps, and other digital platforms for monitoring, data collecting, emotion inference, and feedback translation. Because the data collected by AER systems is the property of the device manufacturers, these spaces are becoming more open, and prone to privacy violation \cite{royakkers2018societal}. Though there are a few traditional privacy limitations in place, it is challenging to ensure privacy when AI-driven inference is involved without suitable monitoring and regulatory mechanisms. Camera-based assistive aids are quickly becoming popular among the sight impaired. AI-based vision technologies and, in certain situations, AER systems are actively used in these assistive technologies. Akter et al.\ \cite{akter2022shared} conducted a couple of surveys on the privacy and ethical considerations associated with these assistive devices. According to their surveys, the majority of respondents were concerned about the fairness, privacy, and other ethical concerns associated with these assistive technologies.

In the last few years, ethical concerns related to privacy have become a promising area of research thanks to the active integration of AI-enabled applications such as camera-based surveillance systems, AER systems, and others. Ribaric et al. \cite{ribaric2016identification} surveyed de-identification techniques for ensuring privacy in vision-based applications such as AER systems and healthcare applications where privacy is critical and provide an insightful discussion on how de-identification can help resolve ethical challenges. Das et al.\ \cite{das2017assisting} provide a procedure for identifying and mitigating privacy-related concerns in camera-based IoT devices in digital homes and other places through privacy-aware notifications and infrastructure. Their work also outlines the technique for privacy-aware video streaming and policy-related guidelines for ensuring privacy and mitigation of the risks involved in vision data (surveillance data, AER systems, etc.) being misused by adversaries. Hunkenschroer et al. \cite{hunkenschroer2022ethics} conducted a systematic review of the literature on ethical problems in AI-based hiring procedures. Though the emphasised problems concern the employment process, some of the issues (human and algorithmic bias, privacy and data leakage hazards, and fairness) are also prevalent in AER systems. Boutros et al. \cite{boutros2022sface} used class conditional generative models to generate a privacy-friendly synthetic faces dataset and trained facial recognition models and tested its performance in three different experimental settings: multi-class classification, label-free knowledge transfer, and combined learning settings. Their results indicate that the synthetic dataset showed a promising performance, and the authors recommend that privacy-friendly synthetic data is good enough to train facial recognition systems.  

\begin{figure}[!ht]
\centering
\includegraphics[width=\columnwidth]{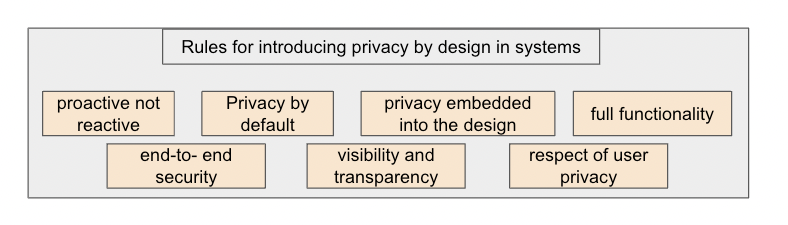}
\caption{Seven rules for introducing privacy by design in systems \cite{cavoukian2009privacy}.  }
\label{fig:privacy}
\end{figure}

Privacy is not just about hiding information; it is about the agency: the agency to opt-in or opt-out. Unfortunately, the concept of agency is frequently overlooked in the design thinking component of AER systems, resulting in biased and untrustworthy AER systems. Because AER systems predict/infer a person's or a social group's emotions, the agency to convey the emotional data (through any input methods such as voice, video, picture, and language) should be with the individual or the social group. Woensel et al. \cite{woensel2019if} raised the problem of agency in AER systems and linked it to data gathering from people and social groupings without proper consent and agency. The critical concern raised in the paper was the potential of using AER systems for targeted and mass surveillance, which in any rational society is considered a violation of social standards, privacy, and ethics. The paper recommended imposing strict controls on data collection for AER systems or for prohibiting them until the necessary ethical standards are satisfied. Cavoukian et al. \cite{cavoukian2009privacy} outlined seven rules for introducing privacy by design in systems. We show these rules in Figure \ref{fig:privacy}. These rules can help improve privacy while providing a reasonable design path toward ethics-centred privacy for AER systems \cite{mohammad2022ethics}.

\subsection{Ethical Concerns Related to Autonomy}
When we adopt AER services in daily life, we willingly cede some of our decision-making power to AI. This may undermine the flourishing of human autonomy with artificial autonomy. Therefore, it is crucial to balance the decision-making power delegated to AER agents and that we retain for ourselves. AER systems must not impair the freedom of their users so they can live according to their standards and norms. 

For AI to yield any benefits for the human race, it must be focused on the autonomy of humans rather than the popular belief of giving more autonomy to machines \cite{prunkl2022human}. This argument stems from the classical discussions on whether AI techniques are tools to help improve life by making tasks easier or AI understanding the problems by itself and fixing them without categorically consenting the humans. Here, it is essential to understand what autonomy means. Autonomy is described as the sense of willingness and a cognitive process of committing to a course of action. Calvo et al. \cite{calvo2020supporting} take a closer look at human autonomy and technology under the pretext of ethics. They highlight that, in 2019, most of the literature around autonomy was focused on machine autonomy, whereas, now, this trend is shifted towards human autonomy-based technology design after critical technical and ethical issues with machine autonomy and design of machine autonomy were highlighted. AER systems are designed to translate the state-of-the-art in human psychology using AI and psycho graphs techniques. Unfortunately, human autonomy and ethical questions such as willingness to interact and adopt are not appropriately addressed. Abbass \cite{abbass2019social}, and \cite{formosa2021robot} argue that since AI techniques are now being integrated into various aspects of society, it is paramount to prefer humans in the loop or humans on the loop-based algorithms for decision making. It will ensure that human autonomy and ethical practices are followed in making critical decisions. 

Emotion recognition systems are trained on the data harvested from social media and digital platforms to understand and infer emotions. Andalibi et al. \cite{andalibi2020human} surveyed 13 social media users about the fact that the data from social media applications are used for training emotion recognition systems without getting users' consent. Even if consent is taken, it is 
collected 
through a `terms and condition' form, which is mainly forced and in a legal language that is not user-friendly.
Their results indicate that most of the participants viewed it as scary, invasive, unethical, and a loss of power and human autonomy. The paper further recommends that ethical usage be ensured in these critical applications at an individual and societal level. Gender bias is another ethical quandary in the AER system, and using these tools in the field necessitates a gender bias evaluation in emotion recognition systems. Domnich et al. \cite{domnich2021responsible} assessed the performance of several AI approaches and showed which kind of networks are employed for certain types of emotions. The results of the experiments revealed that specific AI designs are discriminatory, with significant differences in performance between males and females in terms of emotion recognition.
Another vertical of this discussion on the autonomy-related AER system is the categorisation of complex human emotions into a set of classes and then the offering solutions/interventions based on these categories. Unfortunately, this classification concept has a fundamental weakness since human emotions (both as individuals and as social groupings) are complicated, private, unique, and occasionally indefinable, and reducing these aspects to a data point and using it to tweak the behaviour raises various ethical issues.

\section{Path to a Prosocial Future for AER}

\label{sec:prosocial}

As motivated in the previous sections, AER systems are promising for contributing to social good in a wide variety of applications such as healthcare, education, safety, and law enforcement, but at the same time, it is beset with several risks and perils, which must be addressed. Qadir et al.\  \cite{qadir2022toward} have stressed the need for a more humane human-centred AI that is accountable and have outlined promising directions for achieving accountable human-centred AI. In this section, we highlight some approaches we can adopt to pave the way for a prosocial future for AER systems.

\subsection{Better Awareness and Education}

The development of AER software is bringing enormous changes to society through data analysis. AER Technology has the positive effect of revolutionising many areas by solving various existing problems. On the other hand, AER technologies are two-sided, which can also cause problems. Therefore, it is crucial to raise awareness among the broader population about AER's role in our lives and the use and purchase of AER services. This can help to achieve large-scale adoption of AER services among the general population and minimise the risk of being negatively profiled by AER technology. 

\subsection{Auditable AER}
The auditability of AI describes the possibility of evaluating models, algorithms, and datasets in terms of operation, results, and effects. It has two parts, including technical and ethical. The technical part assesses the reliability and accuracy of results; however, the ethical part apprehends its individual and collective impacts and checks the risks of breaching certain principles, including equality or privacy. AER systems learn from the data they are exposed to and make decisions using ML algorithms. They can develop, or even amplify, biases and discrimination. Therefore, it is essential to audit and test AER algorithms throughout their life cycle to pinpoint the origin of errors and detect risks to avoid their impact on the lives of individuals and society. It will help to systematically probe AER systems, uncover biases, and avoid undesirable behaviour.

\subsection{Explainable and Interpretable AER}

A key reason behind the fragility of AER services is the black-box nature of ML models used for the decision-making process. These ML models are neither explainable nor their outcomes interpretable. To realise the real potential of AER systems, it is highly desirable to make them explainable in a human-understandable way. In recent years, significant research has been devoted to developing novel methods for explaining and interpreting ML models. In the literature, different explainable approaches can be broadly classified as white-box and black-box explanation methods. The white-box explanation method describes the model by identifying the most critical features that contributed to a specific prediction \cite{zeiler2014visualizing}. Another method for white-box explanation is to compute the prediction's gradient concerning individual input samples to discover the prediction's relevant features. White-box explanation mainly provides the model-specific explanation, while the black-box technique provides local explanations of a model for a prediction \cite{lundberg2017unified}. Explaining ML models and their decision is critical, as it is the key enabling factor for building trust and ensuring fairness in decision making. This is also important for AER applications, where decisions directly impact human life.

\subsection{Privacy Preserving AER}
In AER services, the privacy of the users' data is a growing concern, mainly when AER is performed on cloud platforms. AER companies gather a large amount of user data to perform emotion analysis. The data gathered by these companies is kept forever and the user does 
mostly 
not have any 
or little 
control over it. The images, video, and voice samples, 
but also textual bits 
also contain sensitive background information such as faces, gender, language, etc. The leakage of this data can be used maliciously without the user's consent by an eavesdropping adversary and may cause threatening consequences to people's lives. Therefore, it is crucial to utilise privacy-preserving AI models in AER systems to protect users' privacy. The methods and techniques for developing AI systems that ensure privacy falls under the umbrella of privacy-preserving AI. 

Privacy-preserving AI has four major pillars:

\begin{enumerate}
    
    \item \textit{\textbf{Training data privacy}}, which can be ensured by Differentially Private Stochastic Gradient Descent (DPSGD) and PATE \cite{papernot2018scalable} and similar solutions.
    
    \item \textit{\textbf{Input privacy}}, which can be ensured via Homomorphic Encryption, Secure Multiparty Computation (MPC), and Federated Learning.
    
    \item \textit{\textbf{Output privacy}}, which can be ensured by using Homomorphic Encryption, Secure Multiparty Computation (MPC), and Federated Learning
    
     \item  \textit{\textbf{Model privacy}}, which can be ensured by applying differential privacy on the output of an AI model.
\end{enumerate}

\subsection{Ethical Framework for AER}

In recent times, there has been much work on developing ethical principles and frameworks for AI \cite{jobin2019global}. A report on the ethics of AI and the applications of automated emotional intelligence and its risk is presented in \cite{greene2020ethics}. There is a need for similar efforts focused on developing an ethical framework for AI-based AER, which can enable various benefits as presented in Table \ref{prosocial}.



\begin{table}[!ht]
\centering
\caption{Advantages of prosocial AER systems. }
\begin{tabular}{|l|l|l|}
\hline
Ethical principles  & Advantages                                                                                                        \\ \hline
Transparency         & \begin{tabular}[c]{@{}l@{}}Reduces risk\\ Increases fairness\\ Satisfies regulatory and compliance laws\end{tabular} \\ \hline
Full disclosure      & \begin{tabular}[c]{@{}l@{}}Improves explainability\\ Increases understanding\end{tabular}                                                                \\ \hline
Personal consent      & \begin{tabular}[c]{@{}l@{}}Improves reliability and safety\\ Increases regulation \\Protects vulnerable participants\end{tabular}                                                                    \\ \hline
Ethical data sharing & \begin{tabular}[c]{@{}l@{}}Creates an ethical imperative\\ Increases trust \end{tabular}                                                                    \\ \hline
Data ownership       & \begin{tabular}[c]{@{}l@{}}Establishes accountability\\ Assigns responsibility\end{tabular}                                                                \\ \hline
Security and privacy & \begin{tabular}[c]{@{}l@{}}Consent-based data collection\\ Regulated surveillance\\ Improved privacy\end{tabular}                                                                \\ \hline
\end{tabular}
\label{prosocial}
\end{table}

We propose that in order to operate an ethical, privacy-protective AER system, an entity should embrace the following principles:

\begin{itemize}
        \item \textit{\textbf{Transparency}}: An entity must describe its policies related to the duration it retains data, how the data is used, how the government might access the data, and the necessary technical specifications to verify accountability.
        \item \textit{\textbf{Full disclosure}}: An entity must receive informed, written, and specific consent from individuals before enrolling 
        her, him, or them 
        in an AER database. Enrolment is the storage of personal data such as voice and face prints to perform emotion recognition or identification.
        \item \textit{\textbf{Personal consent}}: An entity must receive informed, written consent from an individual before using the individual's data in a manner that was not mentioned in the existing consent. When individuals consent to use an AER system for one purpose, an entity must seek consent from that individual for using AER technology for another purpose. However, users should be free to withdraw their consent at any time. An entity must not use the AER system to determine an individual's colour, race, religion, gender, age, nationality, or disability.
        \item \textit{\textbf{Ethical data sharing}}: Individuals' data should not be shared or sold without the informed, written consent of the individual whose information is being shared or sold.
        \item \textit{\textbf{Data ownership}}: An individual must have the right to access, correct, and remove his or her data print.
        \item \textit{\textbf{Security and privacy}}: AER data must be kept secure and private by the entity maintaining the data. 
\end{itemize}

Simply defining principles is not sufficient. These principles should be embedded into practice and operationalised. An entity must maintain a system that measures compliance with these principles, including an audit trail memorialising the collection, use, and sharing of information in an AER system. The audit trail must include a record of the date, location, consent verification, and provenance of emotional data. It must also allow evaluation of the AER algorithm for accuracy. This data may also be incorporated in a watermark to ease the ability to audit.

\section{Conclusions and Recommendations}
\label{sec:Conclusions}
This paper discussed the promises and perils of artificial intelligence-based automatic emotion recognition systems. We believe AER technology has a wide range of real-life applications; however, we aim to caution AER users and service providers about ethical concerns. AER systems have biases that can lead to incorrect results, just like any other artificial intelligence (AI) based intelligent systems. We cannot 
fully 
rely on AER systems in making decisions; however, help from them can be taken to improve the final decision. We also must carefully consider AER systems' fairness, transparency, accountability, and ethics during their development and applications. For this, we proposed guidelines for designing future prosocial AER solutions. We are summarising below the recommendations for designing such responsible AER systems:
\begin{itemize}

    \item Full examination across various dimensions is required for the data used by AER systems. Expressions of emotion are variable across different languages. This variability must be taken into account while designing datasets, systems, and deployment of AER systems.
    
    \item One needs to examine the choice of AI techniques across interpretability, concerns, privacy, energy efficiency, and data needs. AI tends to perform well for individuals who are well-represented in the data but fails for others. Therefore, it is crucial to explore inclusive methods to avoid spurious correlations that perpetuate sexism,  racism, and stereotypes.
    
    \item AER systems are often trained on static data; however, emotions, perceptions, and behaviour change over time. It is important to incorporate adaptability in AER services for predictions on current data. 
    This may include drifting target learning approaches.
    
    \item Privacy is not only secrecy but also a personal choice. Applying AER to a mass gathering without personal consent is an invasion of privacy, harmful to the individual, and dangerous to society. Therefore, it is important to follow suited privacy principles such as the seven by Cavoukian while designing AER systems. 

    \item It is crucial to realise ethical concerns related to privacy, manipulation, and bias while designing AER systems. Therefore, anonymisation of information at various levels is required.

    \item The use of AER for fully automated decision-making 
    is unsuited. AER systems may be utilised for assistance in decision-making. AER services should be transparent to all stakeholders.
\end{itemize}

These recommendations are primarily for the researchers, engineers, educators, and developers who build, make use of or teach about AER technologies. These guidelines will help engender trust with customers and also improve the profitable drive growth of AER technology. 
With all these guidelines in mind, we shall be ready to fully benefit and enjoy the many good Automatic Emotion Recognition holds as promise.


\end{document}